\documentclass[pra,aps,showpacs,groupedaddress,superscriptaddress,twocolumn]{revtex4-1}

\usepackage[utf8x]{inputenc}
\usepackage{color}
\usepackage{bbm}

\usepackage{nicefrac}
\usepackage{amsfonts,amsmath,amssymb,stmaryrd}
\usepackage{braket}
\usepackage[english]{babel}

\usepackage{tabularx}
\usepackage{multirow} 
\usepackage{array}  
\usepackage{hhline}

\usepackage{graphicx}
\usepackage{subfigure}  
\usepackage{bm}	
\usepackage{bbm}	
\usepackage{hyperref}
\usepackage[pdftex]{epsfig}
\usepackage{mathrsfs}
\usepackage{verbatim}
\usepackage{centernot}

\usepackage{cancel,ifthen}
\usepackage[colorinlistoftodos,prependcaption,textsize=tiny]{todonotes}
\presetkeys{todonotes}{size=\tiny}{}

\InputIfFileExists{gitinfo-latex.inc}{}{}
\usepackage[acronym,nomain]{glossaries}

\usepackage{units}
\usepackage{upgreek}

\renewcommand{\l}{\left(}
\renewcommand{\r}{\right)}

\renewcommand{\H}{\hat{\mathcal{H}}}

\renewcommand{\a}{\ensuremath{\hat{a}}}

\newcommand{\ad}{\hat{a}^\dagger}

\newcommand{\n}{\hat{n}}

\newcommand{\BEC}{\text{BEC}}

\newcommand{\ps}{\hat{\psi}}

\def\mb{\ensuremath{m_\text{B}}}
\def\mi{\ensuremath{m_\text{I}}}
\def\gib{\ensuremath{g_\text{IB}}}

\def\n0{\ensuremath{n_{0}}}

\def\d{\text{d}}
\newcommand{\dirac}[1]{\ensuremath{\delta\left(#1\right)}}
\def\kb{\ensuremath{k_\text{B}}}

\begin{document}
	\bibliographystyle{apsrev4-1}

	\title{Role of thermal two-phonon scattering for impurity dynamics \\
		 in a  low-dimensional BEC}
	
	\author{Tobias Lausch}
	\affiliation{Department of Physics and Research Center OPTIMAS, University of Kaiserslautern, Germany}
	\author{Artur Widera}
	\affiliation{Department of Physics and Research Center OPTIMAS, University of Kaiserslautern, Germany}
	\author{Michael Fleischhauer}
	\affiliation{Department of Physics and Research Center OPTIMAS, University of Kaiserslautern, Germany}
	
	\pacs{...}
	
	\date{\today}
	
	\begin{abstract}
		We numerically study the relaxation dynamics of a single, heavy impurity atom interacting with a finite one- or two-dimensional, ultracold  Bose-gas. While there is a clear separation of time 
		scales between processes resulting from single- and two-phonon scattering in three spatial dimensions, the thermalization in lower dimensions is dominated by two-phonon processes. This is due to infrared divergencies in the corresponding scattering rates in the thermodynamic limit, which are a manifestation of the Mermin-Wagner-Hohenberg theorem.
		It makes it necessary to include second-order phonon scattering in 
		one-dimensional systems even at $T=0$ and above a crossover temperature $T_\textrm{2ph}$ 
		in two spatial dimensions. $T_\textrm{2ph}$ scales inversely with the system size and is much smaller than currently experimentally accessible.
	\end{abstract}
	
	\maketitle

	\section{Introduction}

	Ultracold quantum gases have proven powerful model systems of quantum many-body physics, paving the way for quantum simulations of solid state phenomena or emerging quantum technologies \cite{Bloch2008}. 
	Coupling single or individual impurities via $s-$wave-collisions to a Bose-Einstein condensate (BEC) forms a paradigmatic system of quantum physics. It lies at the heart of many proposals to study and engineer quantum phenomena, including the study of polarons \cite{Mathey2004,Cucchietti2006, Mathy2012,Dehkarghani2015}; application of single atoms for superfluid atomtronics \cite{Micheli2004,Jaksch2005}; and probing of local phase fluctuations \cite{Bruderer2006,Ng2008,Haikka2013}, correlations \cite{Streif2016,Elliott2016} or thermodynamic properties \cite{Sabin2014,Johnson2016}. Extending the model to two impurities, the emergence of quantum correlations via the superfluid was studied \cite{Klein2005,McEndoo2013}.
	%
	
	\begin{figure}[h]
		\includegraphics[width=0.48 \textwidth]{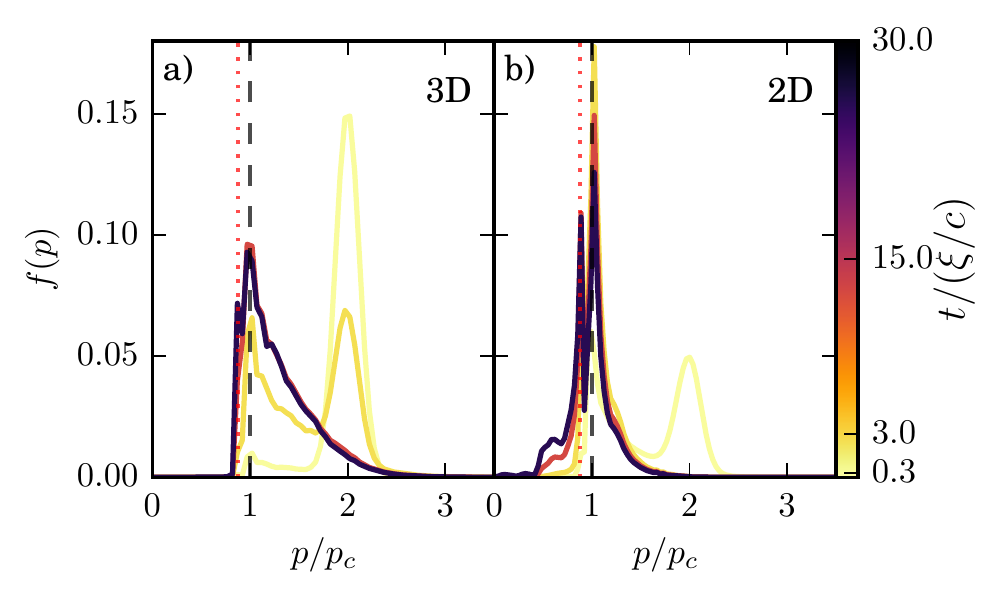}
		\caption{
			Comparison of time evolution (color / grayscale encoded) of impurity momentum distribution centered at $2 \times p_c$ landau critical momentum in 3D (a) and 2D (b) interacting with a BEC at temperature $k_\text{B}T /( c/\xi) = 1$, with $c$ being the speed of sound. Here we assume a peak density of $n_0 \xi^d = 10,~ \gib = 1$, and a mass ratio $m_\text{I}/m_\text{B} = 87 / 39$. The dashed black (dotted red) vertical line indicates a critical momenta $p_c = m_\text{I} c, (p_c^{(1)} = c \sqrt{m_\text{I}^2 - m_\text{B}^2})$ for impurity-BEC interaction.
			While in 3D a separation of time scales between single-phonon scattering and two-phonon scattering events leads to a temporary, prethermalized state, this does not hold in 2D.}
		\label{fig:1}
	\end{figure}

	In recent work, Ref.~\cite{Lausch2017}, we have shown that the cooling dynamics of a single impurity in a three-dimensional (3D) BEC, although the system is non-integrable, features a long-lasting prethermalized stage facilitated by markedly different scattering rates for one-phonon and two-phonon scattering.
	Particularly, for a 3D, weakly interacting Bose-Einstein condensate (BEC) ($n_0\xi^3 \gg 1$), the ratio of scattering rates for spontaneous one-phonon and two-phonon scattering events $\Gamma_\textrm{1ph}^\textrm{sp}$ 
	and $\Gamma_\textrm{2ph}^\textrm{sp}$, respectively, scales as  $\Gamma_\textrm{2ph}^\textrm{sp}/\Gamma_\textrm{1ph}^\textrm{sp}\sim 1/n_0\xi^3 $.
	Here $\xi = 1/\sqrt{2\mb g\n0}$ is the healing length of the \BEC\ , with $\mb$ being the mass of the bosonic atoms, $g$ the interaction strength, and  $\n0$ the BEC density.
	In a superfluid, this leads to a separation of time scales in the relaxation of an initial non-equilibrium state, pointing toward deterministic preparation of a prethermalized state in a non-integrable system. Importantly, as shown in Ref.~\cite{Lausch2017} this behavior is robust for finite temperatures and decreasing interactions, as both single and two-phonon terms are affected. 
	
	It is well known that, for 
	systems with dimension $d\leq 2$, the Mermin-Wagner-Hohenberg (MWH) theorem \cite{Mermin1966,Hohenberg1967} facilitates the creation of long-wavelength excitations at small cost. 
	Thus impurity-superfluid interaction can be easily altered or even dominated by thermal excitations even at small temperatures, leading to qualitatively different dynamics for $d=1,2$ compared to the 3D case. 
	We consider the influence of reduced dimensionality onto the non-equilibrium dynamics of a single impurity in a finite-temperature, low-dimensional BEC, see Fig.\ref{fig:1}a and b.
	In direct comparison to the 3D case, Fig. \ref{fig:1}b, we find that the intermediate regime of prethermalization can no longer be observed in lower dimensions even for very low temperatures $T\neq 0$. We interpret this as a result of the MWH theorem. 
	
	Studying in detail the individual contributions of single- and two-phonon scattering processes to the 
	impurity-quantum gas interaction, we identify to leading order the scattering process realizing the MWH theorem as a two-phonon scattering process. 
	Our findings underline the necessity to include two-phonon processes when studying impurity physics in low-dimensional Bose gases at finite temperatures
	\cite{Wehr2006}, particularly for Bose polarons \cite{Rath2013} in 1D and 2D. Our investigation of impurity dynamics cooled by a low-dimensional superfluid thereby complements the currently emerging description of Bose-polarons beyond the Fr\"ohlich model and its ground state properties in reduced dimensions \cite{Levinsen2017}.
	
	\section{Model and scattering rates}

	We consider the interaction of a single impurity with a trapped, weakly interacting BEC in one and two dimensions at a finite, but very low temperature $T$.
	Although as dictated by the Mermin-Wagner-Hohenberg theorem \cite{Mermin1966,Hohenberg1967} there is no Bose condensation in a homogeneous system,
	condensation can occur in a harmonic trap.
	For an isotropic harmonic potential $V(r)=\mb\omega^2 r^2/2$ with trap frequency $\omega$ the corresponding critical temperatures for an ideal Bose gas are  \cite{MullinAJP2003,Ketterle-PRA-1996}
	\begin{equation}
		\kb\, T_c\Bigr\vert_\textrm{2D} = \omega \left(\frac{N}{\zeta(2)}\right)^{\frac{1}{2}},\quad\kb\, T_c\Bigr\vert_\textrm{1D} = \omega \, \frac{N}{\ln(2N)}.
		\label{eq:T_crit}
	\end{equation}
	Here $N$ is the number of trapped particles and $\zeta$ is the Riemann Zeta function, and we have set $\hbar =1$.
	
	If the condensate atoms are weakly interacting with strength $g$,
	the Bose field can be expressed
	in terms of a macroscopically occupied ground-state wavefunction $\phi(x)$
	and (small) quantum fluctuations, $\ps(x) = \sqrt{N}\phi(x) +\delta\ps(x)$. 
	For a given chemical potential $\mu$, the ground-state wavefunction 
	fulfills the Gross-Pitaevski equation
	\begin{equation}
		H_\textrm{MF} \phi(\vec{x})=\left(-\frac{\Delta}{2\mb} + V(\vec{x})+  g \, n(\vec{x})\right)\phi(\vec{x}) =\mu\phi(\vec{x}),
		\label{eq:GPE}
	\end{equation}
	with $n(\vec{x})=N \vert\phi(\vec{x})\vert^2=N \phi(\vec{x})^2$ being the condensate density. (We have set the particle number in the condensate equal to the total
	particle number, $N$, for simplicity.) A good estimate for the condensate density can be obtained in Thomas Fermi approximation, which gives
	\begin{equation}
		n(\vec{x}) = \frac{\mu -V(\vert\vec{x}\vert)}{ g},\qquad \textrm{for}\enspace \vert \vec{x}\vert \le R.
	\end{equation}
	Here $R$ denotes the Thomas Fermi radius defined by $V(R)=\mb\omega^2 R^2/2 = \mu$. 
	
	Since interactions are assumed to be weak, one can neglect higher than second order terms of the quantum fluctuations $\delta \ps(x)$ in the Hamiltonian. Making use of the Thomas Fermi approximation,
	the grand-canonical Hamiltonian, ${\cal H}=H-\mu N$, expressed in terms of the condensate wave function and quantum fluctuations reads
	\begin{eqnarray}
		{\cal H} &=& H_\textrm{MF} + \int\!\! d^dx\, \, \delta\ps^\dagger(\vec{x})\left(-\frac{\Delta}{2\mb}
		+   g n(\vec{x}) \right)\delta\ps(\vec{x}) \nonumber\\
		&& + \int\!\! d^dx\, \, \frac{ g}{2} n(\vec{x}) \left(\delta\ps^\dagger(\vec{x})\delta\ps^\dagger(\vec{x}) 
		+\delta\ps(\vec{x})\delta\ps(\vec{x})\right).\quad
		\label{eq:H-lin}
	\end{eqnarray}
	The quantum fluctuations and the condensate density can be expanded into plane waves
	assuming periodic 
	boundary conditions with period $L=2 R$. 
	\begin{equation}
		\delta\ps(\vec{x}) = \frac{1}{L^{d/2}} \sum_{\vec{k}} \hat d_{\vec {k}}\, e^{i\vec{k}\cdot\vec{x}},\qquad n(\vec{x}) = \sum_{\vec{q}} n_{\vec{q}} \, e^{i\vec{q}\cdot\vec{x}}
	\end{equation}
	Disregarding the mean-field term  we find for the hamiltonian two contributions ${\cal H}={\cal H}_0+{\cal H}_1$. The first one
	\begin{eqnarray*}
		{\cal H}_0 =  \sum_{\vec{k}} \biggl[\frac{k^2}{2\mb} \hat d_{\vec{k}}^\dagger \hat d_{\vec{k}} +  g  n_0 
		\left(\hat d_{\vec{k}}^\dagger \hat d_{\vec{k}} + \frac{1}{2}\bigl(\hat d_{\vec{k}}^\dagger \hat d_{-\vec{k}}^\dagger+\hat d_{-\vec{k}}\hat d_{\vec{k}}\bigr)\right)\biggr]
	\end{eqnarray*}
	is identical to a homogeneous BEC with condensate density $n_0$ and the second one
	\begin{eqnarray*}
		{\cal H}_1 =  g \, \sum_{\vec{k}\ne \vec{k^\prime}}  n_{\vec{k}-\vec{k}^\prime} 
		\left(\hat d_{\vec{k}}^\dagger \hat d_{\vec{k}^\prime} + \frac{1}{2}\bigl(\hat d_{\vec{k}}^\dagger \hat d_{-\vec{k}^\prime}^\dagger+\hat d_{-\vec{k}}\hat d_{\vec{k}^\prime}\bigr)\right)
	\end{eqnarray*}
	describes the influence of the harmonic confinement.
	${\cal H}_0$ can be diagonalized by a Bogoliubov transformation which relates 
	the boson operators $\hat d_{\vec{k}}$ to plane-wave phonon operators $\hat a_{\vec{k}}$ via
	$\hat d_{\vec {k}} = \cosh\theta_k\, \hat a_{\vec{k}} -\sinh\theta_k\, \hat a_{-\vec{k}}^\dagger$, where
	\begin{equation}
		\begin{array}{c}
			\cosh\theta_k\cr
			\sinh\theta_k
		\end{array} = \frac{1}{\sqrt{2}}\left(\frac{\frac{k^2}{2 \mb}+ g\, n_0}{\omega_k}\pm 1\right)^{1/2}.
	\end{equation}
	This gives 
	\begin{equation}
		{\cal H}_0 = \sum_{\vec{k}} \omega_{\vec{k}} \hat a_{\vec{k}}^\dagger \hat a_{\vec{k}},
	\end{equation}
	with $\omega_{\vec{k}} = c k ( 1 +  k^2 \xi^2  / 2 )^{1/2}$ 
	being the Bogoliubov dispersion relation. Here we have used the speed of sound $c=\sqrt{ g \n0/ \mb}$ 
	corresponding to the
	homogeneous part of the condensate density $n_0$. If the same Bogoliubov transformation is applied to ${\cal H}_1$ one finds a scattering among different momentum modes as well as 
	creation and annihilation of pairs of plane-wave phonons due to the
	influence of the trap potential.
	These couplings are only important, however, for small momenta $k$ on the order of the inverse Thomas Fermi radius $1/R$ or smaller. 
	
	Let us now consider a single impurity of mass $\mi$ that interacts with the BEC atoms with strength $\gib$.  Since 
	${\cal H}_1$ can be considered a perturbation for $k\ge k_c=1/R$,
	we will treat this interaction as in a homogeneous BEC for phonon momenta larger than $k_c$. 
	The impurity-condensate interaction can then be described in terms of plane-wave Bogoliubov phonons for momenta larger than $k_c$ and
	a correction term for smaller momenta. The corresponding Hamiltonian 
	reads  in $d$ spatial dimensions up to a constant mean-field term  \cite{Grusdt2015Varenna}
	\begin{multline}
		\H = \frac{\hat{\vec{p}}^2}{2\mi} + \int \!\! d^d k \left[ \omega_k \ad_{\vec{k}} \a_{\vec{k}} +  \frac{\gib n_0^{1/2}}{(2 \pi)^{d/2}} W_k e^{i \vec{k} \cdot \hat{\vec{r}}} \l \ad_{-\vec{k}} + \a_{\vec{k}}  \r \right]\\ 
		+\frac{\gib}{2(2\pi)^d}  \int_{k_c}\!\! d^dk \, d^dk'  e^{i \l \vec{k}'-\vec{k} \r \cdot \hat{\vec{r}}} \Bigl[ \l W_k W_{k'} + W_k^{-1} W_{k'}^{-1} \r \ad_{\vec{k}} \a_{\vec{k}'} \\
		+ \frac{1}{2} \l W_k W_{k'} - W_k^{-1} W_{k'}^{-1} \r \l \ad_{\vec{k}} \ad_{-\vec{k}'} + \a_{-\vec{k}} \a_{\vec{k}'} \r \Bigr] \\
		+ \H^\textrm{2ph}(k < k_c).\qquad\qquad
		\label{eq:HBogoPolaron}
	\end{multline}
	Here
	\begin{equation}
		W_k =\left[\frac{k^2 \xi^2 }{ 2 + k^2 \xi^2}\right]^{1/4}
	\end{equation} 
	describes the momentum dependence of the coupling of the impurity to Bogoliubov phonons with interaction strength $g_{\rm IB}$.
	Since $d^dk \, W_k \sim dk~k^{d-1}\sqrt{k}$ for $k\to 0$, we have extended the lower limit of the $k$-integral in the first line,
	corresponding to single-phonon terms, from $k_c$ to $0$.
	The first line of eq.(\ref{eq:HBogoPolaron}) is then equivalent to the Fr\"ohlich model studied in condensed matter physics in the context of polaron physics \cite{Froehlich1954}. 
	A simple
	dimensional analysis shows that the two-phonon terms in the second line scale as $1/\sqrt{n_0\xi^d}$ 
	relative to the single-phonon terms,
	and this scaling is sometimes used as an argument to apply the Fr\"ohlich model also to a 
	weakly interacting BEC in 3D, for which $n_0\xi^3\gg 1$, provided the impurity-BEC interaction $\gib$ is sufficiently small \cite{Grusdt2015Varenna}. As will be shown in the following, these arguments do not apply, however, in reduced spatial dimensions, in particular for temperatures $T\ne 0$.
	
	In the above picture a moving impurity can interact with one or more phonons of the \BEC\,  as shown in Figure \ref{fig:2} and thereby resonantly exchange energy spontaneously or induced by thermal phonons. In Born-Markov approximation one can derive a Boltzmann equation for the momentum distribution of the impurity
	$f(\vec{p})$:
	\begin{eqnarray} \label{eq:boltzmann}
		\frac{d f(\vec{p})}{dt} &=& -\left(\Gamma_\textrm{1ph} +\Gamma_\textrm{2ph}\right)  f(\vec{p}) \label{eq:Boltzmann}\\
		&& +\, \,  \textrm{"in" terms}\nonumber
	\end{eqnarray}
	which contains different scattering contributions. The first line of eq.(\ref{eq:Boltzmann}) describes scattering events leading to a reduction of probability density at momentum $\vec{p}$ and
	the second line those that lead to an increase.
	The  term involving the emission of
	a single phonon reads
	\begin{equation} 
		\Gamma_\textrm{1ph}(\vec{p}) = \frac{\gib^2\n0}{(2\pi)^{d-1}} \int \d^d k \, W_k^2 \, \bigl(n_k(T)+1\bigr)\, \dirac{E_{\vec{p},\vec{k}}} 
		\label{eq:thermal_single_phonon}
	\end{equation}
	where $E_{\vec{p},\vec{k}} = \omega_{\vec{k}} - \epsilon_{\vec{p}} + \epsilon_{\vec{p}-\vec{k}}$ denotes the energy difference between in- and outgoing modes,
	with $\epsilon_{\vec{p}} =p^2/2 \mi$ being the kinetic energy of the impurity. 
	
	In addition there are different scattering processes involving two phonons, corresponding to diagrams such as that shown in Fig.\ref{fig:2}b. 
	The rate for the creation of a pair of phonons reads for example
	\begin{eqnarray}\label{eq:two_phonon_spontaneous}
		&&\Gamma_\textrm{2ph}(\vec{p})
		= \frac{g_{\rm IB}^2}{4\pi} \int_{k_c}\!\!\frac{{\rm d}^d k}{(2\pi)^{d-1}}\!\!\int _{k_c}\!\!\frac{{\rm d}^d k^\prime}{(2\pi)^{d-1}} \, \dirac{E_{\vec{p},\vec{k},\vec{k}^\prime}}
		\times\label{eq:two_phonon_thermal}\\
		&& \quad\times \left(\frac{W_{k} W_{k^\prime} -
			W_{k}^{-1} W_{k^\prime}^{-1}}{2}\right)^2  \bigl(n_k(T)+1\bigr)\bigl(n_{k^\prime}(T)+1\bigr)\nonumber 
	\end{eqnarray}
	where $E_{\vec{p},\vec{k},\vec{k}^\prime} = \omega_{\vec{k}} +\omega_{\vec{k}^\prime}- \epsilon_{\vec{p}} + \epsilon_{\vec{p}-\vec{k}-\vec{k}^\prime}$.
	The other scattering terms involving two phonons have an analogous structure. Note that the two-phonon rates are calculated here by taking into account only phonons with momenta
	$k\ge k_c$ and eq. (\ref{eq:two_phonon_thermal}) 
	is thus only a lower bound to the actual rates.
	
	We have shown in Ref.\cite{Lausch2017} that in 3D for a weakly interacting BEC a clear separation of time scales exist between processes resulting from single- and two-phonon scattering.
	For the \textit{spontaneous} processes one finds in a homogeneous condensate
	\begin{equation}
		\frac{\Gamma_\textrm{2ph}^\textrm{sp}}{\Gamma_\textrm{1ph}^\textrm{sp}}\Biggr\vert_\textrm{3D} \sim \frac{1}{n_0\xi^3},
	\end{equation}
	which in a weakly interacting BEC is much less than unity. Consequently in 3D and at $T=0$ single-phonon scattering 
	is much faster than two-phonon scattering.
	Since the rate of \textit{thermal} two-phonon scattering is quadratic in the phonon number, thermal  processes will however
	eventually take over if the temperature is increased.
	The corresponding crossover point $T_\mathrm{2ph}$ can be estimated 
	setting the ratio of one- and two-phonon scattering rates at finite-$T$ equal to unity: $\Gamma_\textrm{2ph}^\textrm{T} / \Gamma_\textrm{1ph}^\textrm{T} \sim \bar{n} /(n_0\xi^3)\bigr\vert_{T_\textrm{2ph}}=1$.
	Here $\bar{n}$ is the thermal phonon number at a characteristic momentum, for which we use $k=1/\xi$.
	This yields apart from factors of order unity
	\begin{equation}
		\frac{T_\textrm{2ph}}{T_c}\Biggr\vert_\textrm{3D} \approx 
		\left(n_0\xi^3\right)^{1/3},
		\label{eq:T2ph_3D}
	\end{equation}
	where we have used the condensation temperature $T_c$ of a homogeneous, non-interacting 3D Bose gas.
	For weak interactions the right hand side of eq.(\ref{eq:T2ph_3D}) is larger than unity and the crossover occurs only above the critical temperature of condensation. 
	
	Thus in a weakly interacting BEC in 3D, single-phonon scattering is much faster than two-phonon scattering for all relevant temperatures. As will be discussed in the following, this behavior changes dramatically in reduced spatial dimensions.

	\section{Scattering of long-wavelength phonons}

	To discuss the relevance of different contributions to the 
	emission of Bolgoliubov phonons in reduced spatial dimensions we express the scattering rates in the form
	$\Gamma_\textrm{1ph} =\int d^d\Omega_k\int\textrm{d}k \, I_\textrm{1ph}(k) \, \delta(E_{\vec{p},\vec{k}})$, and 
	$\Gamma_\textrm{2ph} =\int d^d\Omega_k\int d^d\Omega_{k^\prime}\int \textrm{d}k \int \textrm{d}k^\prime  \,   I_\textrm{2ph}(k,k^\prime)\,  \delta(E_{\vec{p},\vec{k},\vec{k}^\prime})$, where $\int d^d\Omega_k$ denotes angular integration.
	We note that while the factors $W_k$ in the integral kernel $I_\textrm{1ph}(k)$ 
	are well behaved  for $k\to 0$, 
	$I_\textrm{2ph}(k,k^\prime)$ contains also terms
	$W_k^{-1}$ which diverge in this limit. This divergence is compensated in 3D by the $k^2$ term resulting from the density of states, but makes the two-phonon terms relevant in lower dimensions in particular at finite temperatures, since also
	the number of thermal phonons diverges for $k\to 0$.
	
	The integral kernel $I_\textrm{1ph}(k)$ is only weakly dependent on $k$ and is bounded for any finite temperature. This is illustrated in Fig.\ref{fig:2}b and c for the case of an 
	incoming impurity with momentum $p=1.5 p_c$, $p_c=m_I c$ being the Landau critical momentum. This does not hold true, however, for two-phonon scattering terms. The rate for 
	spontaneous two-phonon creation, eq.~(\ref{eq:two_phonon_spontaneous}), for 3D and 2D is bounded, but diverges for $k_c\to0$. Additionally, the thermal two-phonon creation rate converges only in 3D, while in lower dimensions we find a clear divergence in the limit $k_c\to 0$.
	This is illustrated in Figs.{\ref{fig:2}b and c for two-phonon creation.

		\begin{figure}[htb]
			\includegraphics[width=0.49\textwidth]{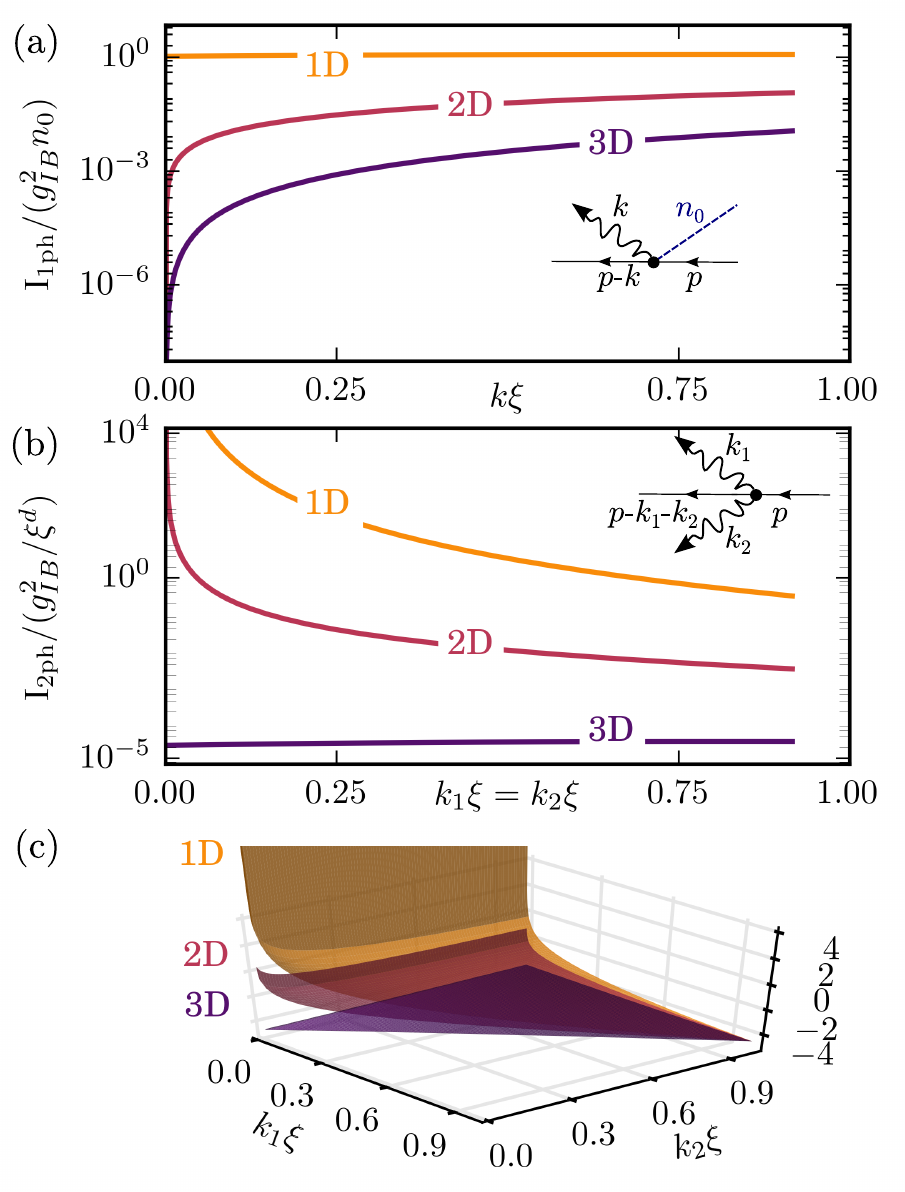}
			\caption{
				Integral kernels of single- and two-phonon thermal scattering rates for an impurity with momentum $p =1.5 p_c$ interacting with a BEC of density $n_0 \xi^d = 5$ in different spatial dimensions for $\kb T/(c/\xi) = 1$.
				Shown are the integral kernels of the creation (a) of a single phonon at momentum $k$.
				The rates for the creation of two phonons with
				momenta $k_1$ and $k_2$ are illustrated in Fig.(c) and Fig.(b) shows a cut for $k_1=k_2$.
			}
			\label{fig:2}
		\end{figure}
		
		This behavior can easily be understood from a simple analysis of the low-momentum scaling of the integral kernels.
		Since $W_k^2\sim k$ and $n_k \sim k^{-1}$ for $k\to 0$ one finds the infra-red scaling 
		summarized in table~\ref{table:1}
		
		\begin{table}[h]
			\begin{tabular}{ | c  | c || c  |  c | c |} 
				\hline 
				&  & $\phantom{\Bigr)}$ \textbf{1D} $\phantom{\Bigr)}$ & \textbf{2D} & \textbf{3D}\\
				\hline
				$\phantom{\Biggr\}} I_\textrm{1ph}(k)\phantom{\Biggr\}}$ & $\phantom{\Biggr\}} T=0\phantom{\Biggr\}} $ 
				& $\phantom{\Biggr\}} k\phantom{\Biggr\}} $ &  $\phantom{\Biggr\}} k^2\phantom{\Biggr\}} $ 
				& $\phantom{\Biggr\}} k^3\phantom{\Biggr\}} $ \\
				\hline
				$\phantom{\Biggr\}} \phantom{\Biggr\}}$ & $\phantom{\Biggr\}} T\ne 0\phantom{\Biggr\}} $ & 
				$\phantom{\Biggr\}} 1\phantom{\Biggr\}} $ &  $\phantom{\Biggr\}} k \phantom{\Biggr\}} $ & 
				$\phantom{\Biggr\}} k^2\phantom{\Biggr\}}  $\\
				\hline
				$\phantom{\Biggr\}} I_\textrm{2ph}(k,k^\prime)\phantom{\Biggr\}}$ &  $\phantom{\Biggr\}} T=0\phantom{\Biggr\}} $ 
				& $\phantom{\Biggr\}}\Bigl(kk^\prime\Bigr)^{-1}\phantom{\Biggr\}}$ &  $\phantom{\Biggr\}} 1\phantom{\Biggr\}} $ & 
				$\phantom{\Biggr\}} kk^\prime\phantom{\Biggr\}} $ \\
				\hline
				$\phantom{\Biggr\}}\phantom{\Biggr\}}$ &  $\phantom{\Biggr\}} T\ne 0\phantom{\Biggr\}} $ 
				& $\phantom{\Biggr\}} \Bigl(kk^\prime\Bigr)^{-2}\phantom{\Biggr\}} $ &  $\phantom{\Biggr\}} 
				\Bigl(kk^\prime\Bigr)^{-1}\phantom{\Biggr\}}  $ 
				& $\phantom{1\Biggr\}} 1\phantom{1\Biggr\}} $ \\	
				\hline
			\end{tabular}
			\caption{Infra-red scaling of integral kernels in scattering rates of spontaneous and thermal single- and two-phonon processes (i.e. for $k, k^\prime\to 0$). }\label{table:1}
		\end{table}

		The infra-red divergences of $I_\textrm{2ph}$ in 1D and  2D is 
		a direct consequence of the Mermin-Wagner-Hohenberg theorem which prevents condensation  in the thermodynamic limit of an infinite homogeneous system.
		Any impurity with momentum $p > p_c$ will induce phonon creation from the condensate at a diverging rate in 1D even at $T=0$ and for any non-vanishing temperature in 2D.
		In order to prevent this "evaporation catastrophe" 
		spatial confinement by a harmonic trap is required, which
		leads to a modification of the low-momentum  density of states as well as a low-momentum cut-off and in this way allows for condensation, however only at low temperatures.

		\section{Cross-over temperature to two-phonon scattering}	
		
		Due to their different temperature scaling,  two-phonon rates
		will exceed the single-phonon ones above a certain temperature $T_\textrm{2ph}$. In the following we will estimate these cross-over temperatures for a trapped 1D and 2D Bose gas. For this it is convenient
		to introduce dimensionless momenta $\kappa=k\xi$  and energies $\epsilon=E/(c/\xi)$ using the characteristic momentum and energy 
		(frequency) scales of a weakly interacting BEC, $1/\xi$ and $c/\xi$. At a given temperature $T$ thermal effects become relevant 
		for $n_k(T)\ge 1$, i.e. for all dimensionless momenta $\kappa$ lower than
		\begin{equation}
			\kappa_\textrm{th} = \frac{k_B T}{c/\xi}=  \frac{T}{T_c} \bigl(n_0 \xi^d\bigr)^{1/d}.
		\end{equation}
		Here we have neglected numerical factors of order unity and have used the condensation temperature in a trap, eq.(\ref{eq:T_crit}), as well
		as the relation between the trap frequency $\omega = \sqrt{2} c/R$, the speed of sound $c$ and the Thomas Fermi radius $R$.
		For experimentally accessible temperatures $T/T_c$ is not very small and thus for  $n_0\xi^d\ge {\cal O}(1)$, thermal
		effects are relevant already for typical phonon momenta on the order of $1/\xi$ and below.
		
		To determine the crossover temperature 
		we compare the two-phonon scattering rates to the single-phonon ones for a characteristic initial
		impurity momentum of $p>p_c = \mi c$ or in dimensionless units $\tilde p = p\xi > \mi c \xi = \mi/ \sqrt{2} \mb\sim {\cal O}(1)$ just above the
		Landau critical momentum $p_c$.


		
		Let us first discuss the single-phonon emission rate, eq.(\ref{eq:thermal_single_phonon}),  in $d=1$ and $d=2$.
		%
		%
		For an impurity momentum $\tilde p$ energy-momentum conservation fixes the allowed phonon momenta to
		%
		$\kappa^2 - 2 \cos\theta \kappa \tilde p + 2\tilde p_c  \kappa \sqrt{1+\kappa^2/2} =0$
		%
		where $\theta$ is the angle between the phonon momentum and the incoming impurity momentum. 
		In one dimension $\cos\theta = \pm 1$.
		For not too large $\tilde p > \tilde p_c$, there is a larger range of
		scattering angles, where $\kappa ={\cal O}(1)$. Thus thermal effects also influence the single phonon scattering for experimentally relevant,
		not too small ratios $T/T_c$. In fact, the single-phonon rate crosses over from a value of 
		\begin{equation}
			\Gamma_\textrm{1ph} \simeq \frac{g_\textrm{IB}^2 n_0}{c (2\pi\xi)^{d-1}}\qquad\quad\textrm{for}\quad  k_B T \ll c/\xi.
			\label{eq:gamma1ph_lowT}
		\end{equation}
		to
		\begin{equation}
			\Gamma_\textrm{1ph} \simeq \frac{g_\textrm{IB}^2 n_0}{c (2\pi \xi)^{d-1}}  \left(\frac{k_B T}{c/\xi}\right)\quad\textrm{for}\quad  k_B T \gg c/\xi.
		\end{equation}

		In Fig.\ref{fig:Gamma_1}a,c  we have plotted the numerically obtained single-phonon rate for $\tilde p=1.5$ as function of $T/T_c$. For very low temperatures
		$\Gamma_\textrm{1ph}$ is independent on $T$ but quickly crosses over into a linear dependence already much below $T_c$.
		
		\begin{figure*}[htb]
			\includegraphics[width=0.8\textwidth]{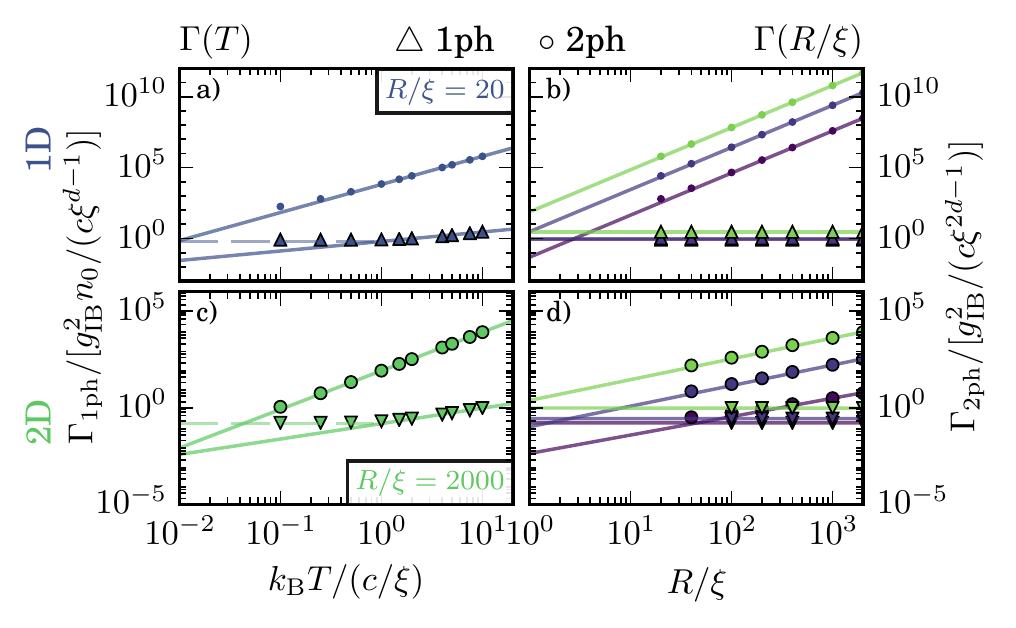}
			\caption{
				Single-phonon $\Gamma_\textrm{1ph}$ ($\triangle$ triangles) and two-phonon $\Gamma_\textrm{2ph}$ ($\circ$ circles) scattering rates as function of temperature in 1D (a) and 2D (c).
				The solid lines are fits to the numerically obtained data points and match the expected behavior.
				The dashed line represents the analytical expectation of 1-phonon scattering for small temperatures, eq.~\ref{eq:gamma1ph_lowT} which is captured by our numerics.
				(b, d) Shows scattering rates for three different temperatures ($k_\text{B}T / (c/\xi) = 10, 2, 0.25$ from top to bottom)  as function of infra-red momentum cut-off identified with the system size $R$ in 1D (b) and in 2D (d), all in log-log scale.
				From our numerical simulations we obtain the theoretically expected scaling with system size (see. eqs.~\ref{eq:Gamma_2-1D-b}, \ref{eq:Gamma_2-2D-b}).
			}
			\label{fig:Gamma_1}
		\end{figure*}
		
		As discussed in the previous section the two-phonon rates are dominated
		by infra-red contributions in 1D and 2D. Thus it is sufficient to replace the corresponding integrand by a low-momentum
		approximation.
		Noting that for $k\to 0$, $W_k^{-2} \approx \sqrt{2}/\kappa$, and $n_k \sim \frac{k_B T}{c/\xi}\frac{1}{\kappa}$ we find in 1D
		\begin{eqnarray}
			\Gamma_\textrm{2ph} ^\textrm{1D}  = \frac{g_\textrm{IB}^2}{16 c}\left(\frac{\kb T}{c/\xi }\right)^2 \!
			\int_{k_c}\!\!\! \textrm{d}k \int_{k_c} \!\!\! \textrm{d}k^\prime \frac{\sqrt{2}}{(k\xi)^2}\frac{\sqrt{2}}{(k^\prime \xi)^2}
			\delta(E_{p,k,k^\prime}).\qquad
			\label{eq:Gamma_2-1D}
		\end{eqnarray}
		Here we have introduced an infra-red cutoff $k_c$ such that the momentum integration extends only over $\vert k \vert > k_c$.
		Energy-momentum conservation $E_{p,k,k^\prime}=0$ gives the dominant contribution for $k^\prime \sim k$. 
		This yields apart from numerical prefactors of order unity
		\begin{eqnarray}
			\Gamma_\textrm{2ph}^\textrm{1D} \sim \frac{g_\textrm{IB}^2}{\xi c} \left(\frac{\kb T}{c/\xi }\right)^2 \,
			\frac{1+6 x^2+x^4}{(1-x^2)^2}\, 
			\left(\frac{R}{\xi}\right)^3
			\label{eq:Gamma_2-1D-b}
		\end{eqnarray}
		Where $x=p/p_c$ and we have identified the infra-red momentum cut-off with the Thomas-Fermi radius $R$ of the trapped BEC.
		When the impurity momentum $p$ approaches the critical momentum $p_c$ the two-phonon rate diverges algebraically, while it quickly approaches a constant value for increasing values of $p>p_c$ as shown in Fig \ref{fig:Gamma_p}a. We find a power-law divergence of $\Gamma_\textrm{2ph}^\textrm{1D} $
		with system size according to $(R/\xi)^3$. The size scaling as 
		well as the temperature dependence of $\Gamma_\textrm{2ph} ^\textrm{1D}$ are verified by numerical calculations of the scattering rates, see
		Fig.\ref{fig:Gamma_1}a,b.
		
		In two spatial dimensions the dominant contributions to the two-phonon scattering rate is given by
		\begin{eqnarray}
			\Gamma_\textrm{2ph} ^\textrm{2D}  &=& \frac{g_\textrm{IB}^2}{16\pi \xi^2}\left(\frac{\kb T}{c/\xi }\right)^2 \!
			\int_{k_c}\!\! \textrm{d}k \int_{k_c} \!\! \textrm{d}k^\prime \frac{\sqrt{2}}{(k\xi)}\frac{\sqrt{2}}{(k^\prime \xi)}\nonumber\\
			&&\times\frac{1}{(2\pi)^2}\int \! d\theta \int \!\d\theta^\prime\, 
			\delta(E_{p,k,k^\prime}).\qquad
			\label{eq:Gamma_2-2D}
		\end{eqnarray}
		Here energy-momentum conservation reads 
		$E_{p,k,k^\prime} = \epsilon_{p,k,k^\prime}/\mi \xi^2=0 $, where $\epsilon_{p,k,k^\prime}\approx \tilde{p}_c (\kappa+\kappa^\prime) -\tilde{p} (\kappa \cos\theta+\kappa^\prime\cos\theta^\prime)$. 
		As will be shown in the Appendix, carrying out the integrations eventually yields
		apart from numerical prefactors of the order of unity, which depend very weakly on system size
		\begin{eqnarray}
			\Gamma_\textrm{2ph}^\textrm{2D} \sim \frac{g_\textrm{IB}^2}{(4\pi)^2\xi^3 c} \left(\frac{\kb T}{c/\xi }\right)^2 \, \left(\frac{R}{\xi}\right).
			\label{eq:Gamma_2-2D-b}
		\end{eqnarray}
		While in 1D there is a strong dependence of the impurity momentum close to the critical value, the scattering rate in 2D is
		only very weakly dependent on $p/p_c$. This is illustrated in Fig.\ref{fig:Gamma_p}, where we have shown the dependence of
		the two-phonon rate in 1D and 2D on the initial impurity momentum for different temperatures.
		We have also verified the scaling of $\Gamma_\textrm{2ph} ^\textrm{2D}$ with system size and 
		temperature by numerical calculations, see
		Fig.\ref{fig:Gamma_1}d.
		
		\begin{figure}[htb]
			\includegraphics[width=0.48\textwidth]{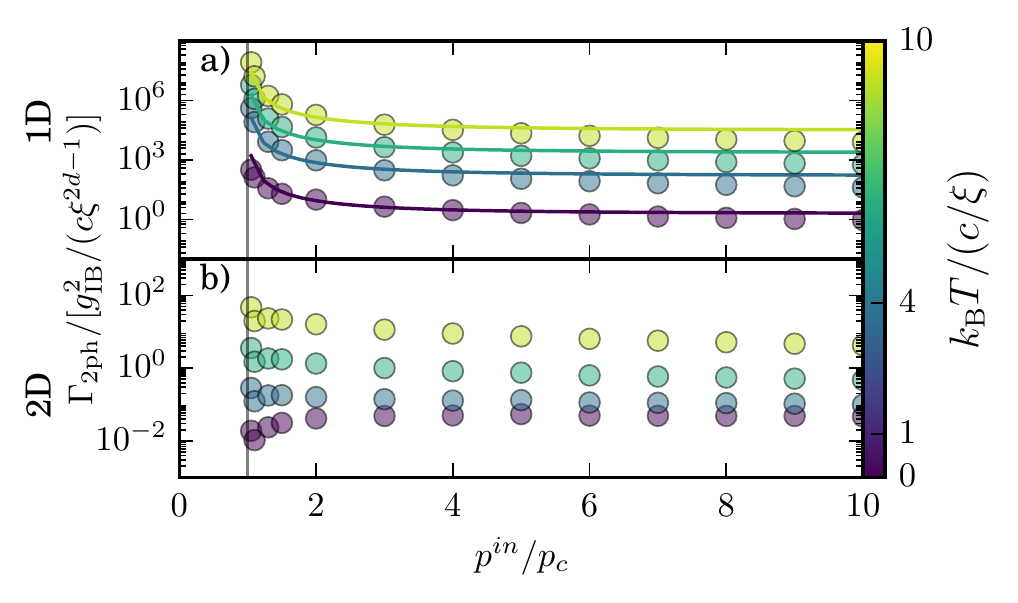}
			\caption{
				Dependence of two-phonon scattering rate in 1D (top) and 2D (bottom) on impurity momentum $p$ for different values of $T$ at $R /\xi = 40$.
				Dots are numerically data, and solid lines are fits according to Eq.~\eqref{eq:Gamma_2-1D-b}.
				In 1D the scattering rate diverges algebraically if $p$ approaches the Landau critical value $p_c$ from above, while in 2D there is only a very weak dependence.
			}
			\label{fig:Gamma_p}
		\end{figure}
		
		We now define the crossover temperature  $T_\textrm{2ph}$ as in the 3D case by setting 
		$\Gamma_\textrm{2ph}^\textrm{T}/\Gamma_\textrm{1ph}^\textrm{T}\Bigr\vert_{T=T_ \textrm{2ph}}=1$.
		To compare the crossover temperatures in 1D and 2D to the critical temperatures of (quasi) condensation, eq.(\ref{eq:T_crit}), 
		we make use of the relation between the trap frequency $\omega$, the Thomas Fermi radius $R$ and the speed of sound $c$: $\omega = 2^{1/d} c/R$. Neglecting numerical prefactors of the order of unity this yields finally 
		\begin{eqnarray}
			\frac{T_\textrm{2ph}}{T_c}\Biggr\vert_\textrm{1D} & \simeq &\ln\Bigl(R/\xi\Bigr) \left(\frac{R}{\xi}\right)^{-3},\label{eq:T2ph-1D}\\
			\frac{T_\textrm{2ph}}{T_c}\Biggr\vert_\textrm{2D} & \simeq & 2\Bigl(n_0\xi^2\Bigr)^{1/2}\left(\frac{R}{\xi}\right)^{-1}.\label{eq:T2ph-2D}\
		\end{eqnarray}
		Here we have used that $\ln(N)\approx \ln(R/\xi)$ since $n_0\xi \sim {\cal{O}}(1)$. Note that  $n_0\xi^2 = l_z/(4 \sqrt{2\pi} a_\textrm{3D})$ in 2D is independent of the particle density due to scale invariance. Here $l_z$ is the transverse
		confinement length and $a_\textrm{3D}$ the 3D scattering length. Thus the prefactor in eq.(\ref{eq:T2ph-2D}) is of order unity. 
		Since typically $R\gg \xi$, the crossover temperatures to two-phonon scattering are much smaller in reduced dimensions as compared to 3D due to the enhanced low-phonon scattering
		and can be substantially below the temperature of condensation.

		\section{Summary and conclusions}

		When an impurity atom interacts with a superfluid Bose Einstein condensate it can induce emission or absorption of single Bogoliubov phonons from or to the condensate.
		Besides these single-phonon processes the impurity can induce scattering of phonons between different modes as well as pair creation or absorption.
		The relative strength of two-phonon processes compared to single-phonon ones is determined by the BEC interaction parameter $1/n_0\xi^d$, the density of phonon states and their thermal occupation probability.
		Since in a weakly interacting 3D condensate $1/n_0\xi^3$ is small, two-phonon processes influence the dynamics of the impurity here only on a much longer time scale than single-phonon scattering, resulting in a separation of time scales and pre-thermalization.
		We have shown that in contrast in lower spatial dimensions and in particular at finite temperatures two-phonon scattering processes play a key role in the impurity-BEC interaction.
		As a consequence a separation of timescales no longer exists. 
		
		In 1D and 2D phonon scattering is dominated by diverging scattering matrix elements involving phonons with momentum 
		$k\to 0$, which  is a manifestation of the Mermin-Wagner-Hohenberg theorem that
		prevents condensation in a homogeneous system. Here BEC formation requires a harmonic confinement which leads to a modified
		density of states at low phonon momenta and 
		to an infra-red cut-off roughly at  the inverse of the Thomas-Fermi radius $R$. The ratio of the  
		characteristic phonon momentum, given by the inverse healing length, to this cut-off value,
		$R/\xi$, defines a second relevant parameter besides the interaction parameter $n_0\xi^d$. Since the thermal occupation of phonon 
		modes increases with decreasing momentum, the relevance of two-phonon scattering is enhanced by temperature effects 
		and there is a crossover between single- and two-phonon dominated regimes at a temperature $T_\textrm{2ph}$.
		Using analytic estimates and 
		numerical simulations we have analyzed the system-size and temperature scalings of two-phonon scattering and have determined
		the cross-over temperatures $T_\textrm{2ph}$. We have shown that 
		in one- and two-dimensional trapped Bose gases $T_\textrm{2ph}$ scales inversely with the system size normalized to the healing length $(R/\xi)$ and can be much below the critical temperature of condensation. Thus two-phonon processes 
		and finite-temperature effects are essential for the impurity dynamics. We expect that thermal effects also play an important role
		in the description of Bose polarons in reduced spatial dimensions 
		and thus must be taken into account.

		\section*{Acknowledgements}

		TL acknowledges financial support from the Carl-Zeiss Stiftung. The work was supported by the Deutsche Forschungsgemeinschaft within the SFB TR49.

		\section*{Appendix}

		In order to derive an analytic estimate for the two-phonon rate in 2D, we 		
		rewrite the Delta-function in eq.(\ref{eq:Gamma_2-2D}) using
		$2\pi \delta(x-y)=\int \textrm{d}\lambda \exp\{i (x-y)\lambda\}$. This allows us to carry out the angular integrations yielding Bessel functions
		\begin{eqnarray}
			&&\frac{1}{(2\pi)^2}\int \! \textrm{d}\theta \!\int \!\textrm{d}\theta^\prime\, 
			\delta(E_{p,k,k^\prime}) = \frac{\mi\xi^2}{(2\pi)^3} \times \nonumber\\
			&&\qquad\times 
			\int \textrm{d}\lambda \,	e^{-i\tilde{p}_c(\kappa+\kappa^\prime)\lambda}\!
			\int \! \textrm{d}\theta \, e^{i\lambda\tilde{p}\kappa\cos\theta }\int \! \textrm{d}\theta^\prime\,  e^{i\lambda\tilde{p}\kappa^\prime\cos\theta^\prime}\nonumber\\
			&& \quad
			= \frac{\mi\xi^2}{2\pi} \int \textrm{d}\lambda \,	e^{-i\tilde{p}_c(\kappa+\kappa^\prime)\lambda}		
			J_0\Bigl(\vert \tilde{p} \lambda \kappa\vert\Bigr)J_0\Bigl(\vert \tilde{p} \lambda \kappa^\prime\vert\Bigr)
			\label{eq:Bessel}	
		\end{eqnarray}
		Note that $\tilde{p}>\tilde{p}_c = \mi/(\sqrt{2}\mb)\sim {\cal O}(1)$. Since the most dominant contribution to the $(\kappa,\kappa^\prime)$-integrals comes from small values of $\kappa$ and $\kappa^\prime$, one finds that the integral in the last line 
		in eq.(\ref{eq:Bessel})
		can be estimated as $C/(\kappa +\kappa^\prime)$, with some numerical constant $C$ of order unity. This then yields for the two-phonon rate
		\begin{eqnarray}
			\Gamma_\textrm{2ph} ^\textrm{2D}  = \frac{g_\textrm{IB}^2\tilde{p}_c}{(4\pi)^2\xi^3c}\left(\frac{\kb T}{c/\xi }\right)^2 \!
			\int_{\kappa_c}\!\! \textrm{d}\kappa \int_{\kappa_c} \!\! \textrm{d}\kappa^\prime \frac{C}{\kappa\kappa^\prime(\kappa+\kappa^\prime)}.	\qquad
		\end{eqnarray}	
		Carrying out the $\kappa$ and $\kappa^\prime$ integration finally leads to eq.(\ref{eq:Gamma_2-2D}).
		\bibliography{bibliography}
	\end{document}